\documentclass[
 amsmath,
 amssymb,
 aps,
 prb,
 reprint,
 groupedaddress
]{revtex4-2}

\usepackage{graphicx}
\usepackage{dcolumn}
\usepackage{bm}
\usepackage{chemformula} 
\usepackage[T1]{fontenc} 
\usepackage{array}
\usepackage{amssymb}

\begin{document}


\title{
  Anomalous enhancement of large-momentum scattering by electron-electron interaction in moir\'e superlattices
}


\author{Taiki Sato} \email{taiki-sato@stat.phys.titech.ac.jp}
\author{Hiroaki Ishizuka} \email{ishizuka@hiroishizuka.com}
\affiliation{Department of Physics, Institute of Science Tokyo, Meguro, Tokyo 152-8551.}



\date{\today}

\begin{abstract}
Using a microscopic model, we show that the large-momentum scattering of electrons in flat bands deviates significantly from that of the Coulomb scattering.
In particular, we find that large-momentum scattering is enhanced at $\theta\lesssim4^\circ$, with a non-monotonic momentum dependence appearing near the magic angle.
For $\theta \gtrsim 1.2^\circ$, the enhanced large-momentum scattering can be attributed to the compact Wannier function.
On the other hand, for $\theta\lesssim1.2^\circ$, the nonmonotonic momentum dependence of the interaction matrix cannot be explained by a simple Wannier orbital, indicating a nontrivial modification of the el-el scattering.
Notably, the range of angles $\theta$ where the large-momentum scattering is enhanced differs from the magic angles at which nearly-flat bands emerge, suggesting that the angle dependence of material properties provides information about the effect of interaction.
The results highlight unusual features of the interaction in moir\'e graphene.
\end{abstract}

\maketitle

\section{Introduction}

\begin{figure}
    \centering
    \includegraphics[width=\linewidth]{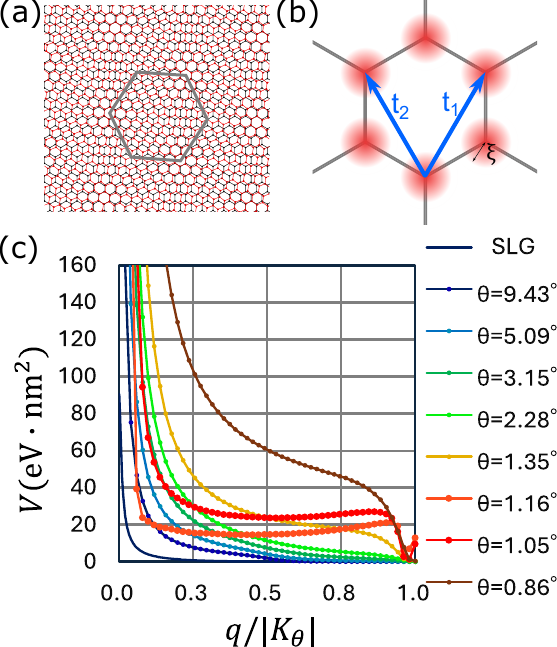}
    \caption{Schematic of a moir\'e structure and the momentum dependence of electron-electron interaction matrix. Schematics of (a) the lattice structure of twisted bilayer graphene for $\theta=7.34^\circ$, and (b) a tight-binding model taking account of small Wannier orbital radius. (c) The $q$ dependence of the interaction matrix for the flat bands and the interaction matrix for pristine graphene along path $A$. The path along which the interaction matrices were calculated is given in Fig.~\ref{fig:Fig2}(a).
    }
    \label{fig:Fig1}
\end{figure}

Electron-electron interactions give rise to rich phases of matter, including Mott insulating, superconducting, and magnetic states.
A compelling realization of such a correlated system is twisted van der Waals heterostructures [Fig.\ref{fig:Fig1}(a)], in which the high tunability of electronic bands allows realizing flat bands that relatively enhance the electron-electron (el-el) interaction~\cite{dSantos2007a,Mele2010a,Shallcross2010a,Suarez2010a,Bistritzer2010a,Bistritzer2011a, TdLaissardiere2010a}.
The existence of these narrow electronic bands, initially predicted theoretically, has since been confirmed experimentally~\cite{Li2010a, Utama2021a, Lisi2021a}.
Furthermore, recent experiments have revealed a variety of correlated phases in these systems, including superconductivity~\cite{Cao2018a,Cao2018b,Lu2019a,Yankowitz2019a, Arora2020a}, ferromagnetism~\cite{Aaron2019a, Serlin2020a}, integer and fractional Chern insulating phases~\cite{Nuckolls2020a, Choi2021a, Xie2021a}, nematicity~\cite{Choi2019a,Kerelsky2019a, Cao2021a}, and charge order~\cite{Jiang2019a}. These discoveries highlight twisted moir\'e systems as a promising platform for studying and engineering electron correlations.

The behavior of strongly correlated systems is generally sensitive to both the form and strength of the effective el-el interaction.
In moir\'e graphene, theoretical studies have suggested that el-el interactions play a significant role in shaping electronic states, even at the mean-field level~\cite{Guinea2018a,Gonzalez2021a,JimenoPozo2023a}, and
that a heavy-fermion-like description may be relevant for the flat bands~\cite{Song2022a,Chou2023a}.
Beyond the mean-field theory, sophisticated numerical methods have provided further insights into the correlation effects in these systems~\cite{Rai2024a,Xiao2024a}, reinforcing the idea that electronic states and their dynamics are highly sensitive to the el-el interaction.
Experimentally, a large $T$-linear resistivity has been observed over a range of twist angles, including those away from the magic angle~\cite{Chung2018a, Polshyn2019a, Cao2020a}, while a large $T^2$-resistivity has been reported near full filling~\cite{Jaoui2022a}.
These transport behaviors have been attributed to enhanced electron-phonon and el-el interactions~\cite{Ishizuka2021a,Ishizuka2022a,Mahew2024a}.
The pronounced sensitivity of moir\'e graphene to the nature of interaction provides a unique opportunity to explore the relationship between correlated phases and the nature of el-el interaction.

In this work, we investigate the el-el interaction matrix in moir\'e graphene using a microscopic theory based on a tight-binding model~\cite{TdLaissardiere2010a, TdLaissardiere2012a}.
We show that, beyond its magnitude, the structure of the el-el interaction matrix for the flat bands undergoes significant changes with the twist angle [Fig.~\ref{fig:Fig1}(c)].
Specifically, we find that large-momentum scattering is enhanced compared to the bare Coulomb interaction.
The enhancement of large-momentum scattering at twist angles $\theta \lesssim 1.2^\circ$ shows a nonmonotonic $q$ dependence, which is qualitatively different from the typical cases. 
For the cases $\theta\gtrsim1.3^\circ$, the interaction matrix has a form qualitatively similar to the usual Coulomb interaction.
However, a considerable increase of the large-$q$ scattering occurs in a wide range of twist angles up to a few degrees.
Our results indicate that, in addition to the emergence of narrow electronic bands, moir\'e graphene hosts a highly nontrivial el-el scattering that is strongly influenced by the underlying Wannier orbitals, with potential implications for its correlated phases.


\section{Model and Method}

\subsection{Tight-binding model}

To study the effective interaction matrix at different twist angles, we calculate the electronic bands of moir\'e graphene using a tight-binding model~\cite{TdLaissardiere2010a}, focusing on commensurate twist angles $\theta$. 
Moir\`e graphene consists of two layers of graphene, labeled 1 and 2, with layer 2 twisted by an angle $\theta$ with respect to layer 1.

The system is spatially periodic for certain twist angles, which are called commensurate twist angles, allowing us to apply Bloch's theorem.
The primitive vectors $\mathbf{t}_1, \mathbf{t}_2$ of the periodic moir\'e lattice are characterized by two integers $(n, m)$ ,
$\mathbf{t}_1 = -m \mathbf{a}_1 + (n+m) \mathbf{a}_2, ~\mathbf{t}_2 = (n+m) \mathbf{a}_1 -  n \mathbf{a}_2.$
Here, $\mathbf{a}_1=(a_0,0)$ and $\mathbf{a}_2 = ( a_0/2, \sqrt{3}a_0/2 )$ ($a_0=2.46$ \AA) are the primitive vectors of the pristine graphene in layer 1.
The twist angle of this model is $\theta = \arccos\left(\frac{n^2+4nm+m^2}{2(n^2 + nm + m^2)}\right)$ and the reciprocal lattice vectors are
$\mathbf{G}_1 = \frac{1}{N_c}(n \mathbf{g}_1 + (n+m)\mathbf{g}_2 ),~\mathbf{G}_1 = \frac{1}{N_c}((n+m) \mathbf{g}_1 + m\mathbf{g}_2 ),$
where $4N_c=4(n^2 + nm + m^2)$ is the number of carbon atoms in the unit cell and $\mathbf{g}_i$ are the reciprocal vectors of pristine graphene in layer 1; $\mathbf{a}_i \cdot \mathbf{g}_j = 2\pi\delta_{ij}$.
To avoid confusion, we call the Brillouin zone of moir\'e superlattice the moir\'e Brillouin zone. The position of two $K$ points in the moir\'e Brillouin zone reads $\mathbf{K}_1 = \frac{1}{3}(\mathbf{G}_1+\mathbf{G}_2),~ \mathbf{K}_2 =\frac{1}{3}(-\mathbf{G}_1+2\mathbf{G}_2).$
We note that the length of primitive vectors $|t_{\theta}|=|\mathbf{t}_{1}|=|\mathbf{t}_{2}|$ and position vector of the K-points $|K_\theta|=|\mathbf{K}_1|=|\mathbf{K}_2|$ depends on the twist angle.

To calculate the el-el interaction matrix, we consider a Hamiltonian $H = H_0 + H_I,$ where $H_0$ and $H_I$ are the single-particle and interaction terms of the Hamiltonian, respectively.
The single-particle Hamiltonian $H_0$ is based on a transferrable tight-binding model for carbon atoms~\cite{Xu1992a, TdLaissardiere2010a}, which we explained in the Supplemental Material~\cite{suppl} (see also references~\cite{Xu1992a, TdLaissardiere2010a, TdLaissardiere2012a, Slater1954a} therein). The relaxation of atoms was not considered in the calculation below as its impact on electron bands is reported to be relatively small for twist angles greater than the first magic angle~\cite{Carr2019a}.

\subsection{Interaction matrix}

To investigate the nature of the el-el scattering, we calculated the interaction matrix for the flat bands.
To this end, we consider the Yukawa interaction between the electrons,
\begin{equation}\begin{split}
    &H_I =\\ &\frac{1}{2} \sum_{\sigma, \sigma'}
    \int \mathrm{d}^3 r \mathrm{d}^3 r' \hat{\psi}^\dagger (\mathbf{r}, \sigma) 
    \hat{\psi} (\mathbf{r}, \sigma) 
    V(\mathbf{r} - \mathbf{r}')
    \hat{\psi}^\dagger (\mathbf{r}', \sigma')
    \hat{\psi} (\mathbf{r}', \sigma').  \label{interaction-hamiltonian}
\end{split}\end{equation}
Here, $\hat{\psi}(\mathbf{r}, \sigma)$ is the field operator for electrons at position $\mathbf{r}$ and spin $\sigma$, and $V(\mathbf{r})=\frac{e^2}{4 \pi \epsilon_0}\frac{\exp(- q_0 |\mathbf{r}|)}{|\mathbf{r}|}$ with $e<0$ being the electron charge, $\epsilon_0$ is the permittivity of vacuum, and $q_0^{-1}$ is the screening length. The interaction $V(\mathbf{r})$ is equivalent to the Coulomb interaction if $q_0=0$. In the following, we take $q_0^{-1}= 10000$ \AA.

First, we write the el-el interaction Hamiltonian in Eq.~\eqref{interaction-hamiltonian} using the density operator, 
\begin{equation}\begin{split}
    H_I &= \frac{1}{2S} \sum_{\mathbf{q}} \tilde{V}(\mathbf{q} ) \tilde{\rho}(-\mathbf{q} ) \tilde{\rho}(\mathbf{q} ),
\end{split}\end{equation}
where $\tilde{\rho}(\mathbf{q} )$ and $\tilde{V}(\mathbf{q})$ are the Fourier transform of the density operator $\hat{\rho}(\mathbf{r}) = \sum_{\sigma} \hat{\psi}^\dagger(\mathbf{r},\sigma)  \hat{\psi}(\mathbf{r},\sigma) $ and $V(\mathbf{r})$, respectively, $\mathbf{q}$ is the momentum, and $S$ is the system size.
Using the annihilation (creation) operator for the electrons in $n$th band with momentum $\mathbf{k}$, $\hat{c}_{n, \mathbf{k}}$ ($\hat{c}_{n, \mathbf{k}}^\dagger$), and its eigenstate wavefunction $\phi_{n, \mathbf{k}}$, calculated using a tight-binding model~\cite{TdLaissardiere2010a},
the density operator in the momentum space reads
\begin{equation}\begin{split}
    \tilde{\rho} (\mathbf{q}) &= \sum_{n, m, \mathbf{k}}  \big<  \phi_{n, \overline{\mathbf{k}-\mathbf{q}} } \big| e^{-i\mathbf{q}\cdot\hat{\mathbf{r}} } \big|  \phi_{m, \mathbf{k} }  \big>   \hat{c}_{n, \overline{\mathbf{k}-\mathbf{q}} }^\dagger  \hat{c}_{m, \mathbf{k} },\label{eq:rho_q}
\end{split}\end{equation}
where $\overline{\mathbf{k}}$ represents the momentum vector in the first moir\'e Brillouin zone that is equivalent to $\mathbf{k}$, and $\hat{\mathbf{r}}$ is the position operator. 
Using Eq.~\eqref{eq:rho_q}, the interaction Hamiltonian reads
\begin{equation}\begin{split}
    H_I &= \frac{1}{2S} \sum_{ \substack{n, n', m, m'\\ \mathbf{k}, \mathbf{k}', \mathbf{q} } }
    V_{ \mathbf{k}, \mathbf{k}', \mathbf{q} }^{n, n', m, m'} \hat{c}_{n', \overline{\mathbf{k'+q}} }^\dagger \hat{c}_{m', \mathbf{k}'} \hat{c}_{n, \overline{\mathbf{k-q}} }^\dagger \hat{c}_{m, \mathbf{k} }, \label{interaction-hamiltonian-2}
\end{split}\end{equation}
where
\begin{equation}\begin{split}
    V_{ \mathbf{k}, \mathbf{k}', \mathbf{q} }^{n, n', m, m'} =
     \sum_{ \overline{\mathbf{q}'} = \mathbf{q} } \tilde{V}(\mathbf{q}' ) \big<  \phi_{n', \overline{\mathbf{k}'+\mathbf{q}'} } \big| e^{i\mathbf{q}'\cdot\hat{\mathbf{r}} } \big|  \phi_{m', \mathbf{k}' }  \big>\\ \times\big<  \phi_{n, \overline{\mathbf{k}-\mathbf{q}'} } \big| e^{-i\mathbf{q}'\cdot\hat{\mathbf{r}} } \big|  \phi_{m, \mathbf{k} }  \big>,   \label{interaction-matrix}
\end{split}\end{equation}   
is the interaction matrix. The interaction matrix corresponds to electron-electron scattering.
In Eq.~\eqref{interaction-matrix}, the sum is taken over all momenta $\mathbf{q}'$ which are equivalent to $\mathbf{q}$ in the first moir\'e Brillouin zone.

To compute Eq~\eqref{interaction-matrix}, we assumed that the $p_z$ orbitals of carbon atoms are well localized on each site.
In the momentum space representation, the eigenstate of $H_0$ reads
\\$\big| \phi_{n, \mathbf{k}} \big> = \sum_{\mathbf{G},\alpha,l}  \phi_{n, \mathbf{k}}(\mathbf{G}, \alpha, l) \big| \mathbf{k} + \mathbf{G}, \alpha, l \big> $, where $\mathbf{G}$ denotes reciprocal lattice vectors, $\big| \mathbf{k}, \alpha, l \big> = \sqrt{\frac{4}{N_{\mathrm{atom}}} } \sum_{n} \exp(i \mathbf{k}\cdot \mathbf{r}_{n,\alpha} ) \big| p_z(\mathbf{r}-\mathbf{r}_{n, \alpha, l} ) \big> $, and $\big| p_z(\mathbf{r}-\mathbf{r}_{n, \alpha, l}) \big>$ is the wavefunction of the $p_z$ orbital on the $\alpha$th sublattice in the $n$th unit cell of the layer $l$, and $N_{\mathrm{atom}}$ represents the number of carbon atoms. For the localized $p_z$ orbitals, i.e.
$\big< p_z(\mathbf{r}-\mathbf{r}_{n,\alpha, l}) \big| e^{-i\mathbf{q}\cdot \hat{\mathbf{r}} } \big|p_z(\mathbf{r}-\mathbf{r}_{m,\beta, l'})\big> \approx \delta_{nm}\delta_{\alpha\beta} \delta_{ll'}\big< p_z(\mathbf{r}-\mathbf{r}_{n,\alpha, l}) \big| e^{-i\mathbf{q}\cdot  \hat{\mathbf{r}} } \big|p_z(\mathbf{r}-\mathbf{r}_{n,\alpha, l})\big> $, 
the brackets in Eq.~\eqref{interaction-matrix} become
\begin{equation}\begin{split}
    \big<  \phi_{n, \overline{\mathbf{k}-\mathbf{q}} } \big| e^{-i\mathbf{q}\cdot\hat{\mathbf{r}} } \big|  \phi_{m, \mathbf{k} }  \big>  
    =& \big< p_z (\mathbf{r}) \big|  e^{-i\mathbf{q}\cdot\hat{\mathbf{r}} } \big|  p_z (\mathbf{r}) \big> \\
    &\times\sum_{\substack{\mathbf{G}, \alpha, l }} \phi_{m, \mathbf{k} } (\mathbf{G-G_{k-q}}+\mathbf{g}^{(l)}, \alpha, l) \\
    &\times \phi_{n, \overline{\mathbf{k}-\mathbf{q}} }^* (\mathbf{G}, \alpha, l)  e^{i\mathbf{g}^{(l)}\cdot\delta_\alpha^{(l)}},
\end{split}\end{equation}
where $\mathbf{g}^{(l)}$ is the reciprocal vector of the $l$th layer in the moir\'e graphene such that the wave number $\mathbf{G-G_{k-q}}+\mathbf{g}^{(l)}$ is in the first Brillouin zone of pristine graphene in layer $l$, and $\mathbf{G_k} = \mathbf{k} - \overline{\mathbf{k}}$ is a vector on the reciprocal lattice that relates $\mathbf{k}$ and the equivalent wavenumber in the first moir\'e Brillouin zone $\overline{\mathbf{k}}$.

Compared to recent studies considering Hubbard-like interactions and near-neighbor interactions~\cite{Calderon2020a,Song2022a}, the current method allows investigating the momentum dependence in detail by taking into account interactions between orbitals at a distance.
In fact, recent work shows a slow decay of the interaction matrix with respect to distance~\cite{Calderon2020a}, implying that further neighbor interactions may considerably affect the momentum dependence.%

\begin{figure}
    \centering
    \includegraphics[]{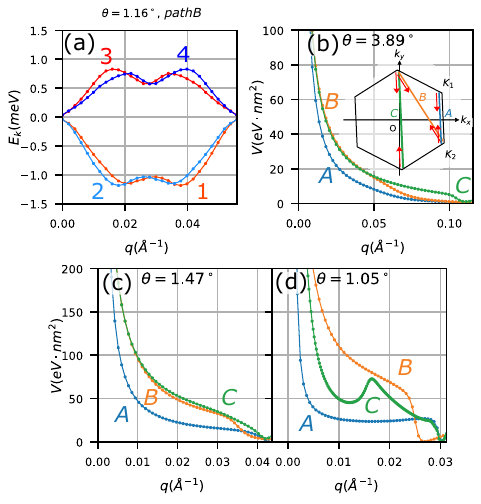}
    \caption{The band and interaction matrix of flat-band electrons. (a) The dispersion $E_{\mathbf{k}}$ of flat bands $n=1,2,3,4$ for twist angle $\theta = 1.16^\circ$ calculated along the path $B$ (see inset of (b)). The even and odd bands belong to different valleys. The interaction matrix for twist angles (b) $\theta=3.89^\circ$, (c) $1.47^\circ$, and (d) $1.05^\circ$. Inset of (b) shows three paths along which the interaction matrix was calculated: (A) $\mathbf{k}=29/30\mathbf{K}_1+1/30\mathbf{K}_2$ and $\mathbf{k}'=1/30\mathbf{K}_1+29/30\mathbf{K}_2$, (B) $\mathbf{k}=29/30\mathbf{K}_1-28/30\mathbf{K}_2$ and $\mathbf{k}'=1/30\mathbf{K}_1+28/30\mathbf{K}_2$, and (C) $\mathbf{k}=29/30\mathbf{K}_1-28/30\mathbf{K}_2$ and $\mathbf{k}'=-28/30\mathbf{K}_1+29/30\mathbf{K}_2$. In all cases, the direction of $\mathbf{q}$ is taken so that $\overline{\mathbf{k-q}}$ moves from $\mathbf{k}$ to $\mathbf{k'}$.}
    \label{fig:Fig2}
\end{figure}

\section{Results}
Typical $\mathbf{q}$ dependence of $|V_{ \mathbf{k}, \mathbf{k}', \mathbf{q} }^{n, n', m, m'}|$ for the flat bands 
is plotted in Fig.~\ref{fig:Fig2}(b) and \ref{fig:Fig2}(c).
The flat bands consist of four conduction and valence bands, each of which corresponds to different spins and valley degrees of freedom\cite{Koshino2018a,Kang2018a,Carr2019a}.
Focusing on one of the two spin sectors, we label the two valence bands $n=1,2$, and the conduction bands $n=3,4$; the valence (conduction) bands $n=1$ and $2$ ($n=3$ and $4$) belong to different valleys of pristine graphene, as shown in Fig.~\ref{fig:Fig2}(a); it is essentially the same as in the one reported in previous works~\cite{TdLaissardiere2010a, TdLaissardiere2012a}.
Figures~\ref{fig:Fig2}(b)-\ref{fig:Fig2}(d) shows the result of $|V_{ \mathbf{k}, \mathbf{k}', \mathbf{q} }^{3,4,3,4}|$ along the paths in the inset of Fig.~\ref{fig:Fig2}(b).
In the figures, $\mathbf{k}$ and $\mathbf{k}'$ are fixed to the wavenumbers corresponding to the two ends of the path, and the direction of $\mathbf{q}$ is along the path as shown in Fig.~\ref{fig:Fig2}(b).
Figures~\ref{fig:Fig2}(b) and \ref{fig:Fig2}(c) shows that, for most $\theta$, $|V_{ \mathbf{k}, \mathbf{k}', \mathbf{q} }^{3,4,3,4}|$ decays monotonically with increasing $q = |\mathbf{q}|$, similar to that of the Coulomb interaction.
Among the three results, the amplitude of $|V_{ \mathbf{k}, \mathbf{k}', \mathbf{q} }^{3,4,3,4}|$ for path $A$ is smaller than those for paths $B$ and $C$.
This suppression resembles the suppression of backscattering in the Dirac electrons.
The results are similar for other $|V_{ \mathbf{k}, \mathbf{k}', \mathbf{q} }^{n, n', m, m'}|$, as shown in the Supplemental Material~\cite{suppl}.
Additionally, results are similar for another symmetrical path that connects the M point to the $\Gamma$ point, which is also shown in the Supplemental Material~\cite{suppl}.%

On the other hand, $V_{ \mathbf{k}, \mathbf{k}', \mathbf{q} }^{n, n', m, m'}$ along paths $A$ and $C$ show nonmonotonic $q$ dependence near the magic angle, as seen in the results for $\theta=1.05^\circ$ in Fig.~\ref{fig:Fig2}(d).
The results show that the large angular momentum scattering by the el-el interaction is enhanced near the magic angle, indicating a unique feature of the el-el scattering in moir\'e graphene.
The enhancement of large momentum scattering is consistent with what is predicted in a previous work~\cite{Ishizuka2022a}.
However, the nonmonotonic $q$ dependence was not predicted.
As shown in Fig.~\ref{fig:Fig1}(c), the enhanced large-momentum scattering is seen in a range of angles around $\theta=1.1^\circ$.


For further analysis, we compare the results in Fig.~\ref{fig:Fig2} to an interaction matrix of the form
\begin{equation}\begin{split}
    V_{ \mathbf{k}, \mathbf{k}', \mathbf{q} }^{a,b,c,d} =& e^{ i( \mathbf{k}-\overline{\mathbf{k-q}} + \mathbf{k}'-\overline{\mathbf{k'+q}} ) \cdot \Delta_1 } \\
    \times\sum_{\overline{\mathbf{q}'} = \mathbf{q} }& \tilde{V}(\mathbf{q}') |\eta_{\mathbf{q}'}|^2 \frac{g_{\mathbf{k-q'}}^* g_{\mathbf{k}} + ac }{2} \frac{ g_{\mathbf{k'+q'}}^* g_{\mathbf{k}'} + bd }{2}, \label{eq:interaction-matrix-fit}
\end{split}\end{equation}
where $\Delta_i$ is a vector connecting the three nearest neighbor sites in a honeycomb lattice,
$|\eta_{\mathbf q}|^2= e^{-\frac{q^2 \xi_G^2}{2} }$
with $\xi_G$ being a fitting parameter, $g_{\mathbf{k}} = \sum_{i=1}^3 e^{i\mathbf{k} \cdot \Delta_i}/\left|\sum_{i=1}^3 e^{i\mathbf{k} \cdot \Delta_i}
\right|$, $\tilde{V}(\mathbf{q}) = \frac{e^2}{2\epsilon} \frac{1}{\sqrt{q_{\mathrm{TF}}^2 + \mathbf{q}^2 }}$ is the Yukawa interaction with $\epsilon$ being the electrical permittivity and $q_{\mathrm{TF}}^{-1}$ is the Thomas-Fermi screening length.
The band indices $a,b,c,d=\pm 1$ are left merely to show which matrix element we fitted.

The interaction matrix in Eq.~\eqref{eq:interaction-matrix-fit} corresponds to that of a tight-binding model on a honeycomb lattice [Fig. \ref{fig:Fig1}(b)].
Consider a single-particle Hamiltonian
\begin{equation}
    \tilde{H}_0 = \sum_{\mathbf{k}} \big(\hat{\psi}_{\mathbf{k}, 1}^\dagger, \hat{\psi}_{\mathbf{k}, 2}^\dagger \big) 
    \left(
    \begin{array}{cc}
    0 & h_{\mathbf{k}} \\
    h_{\mathbf{k}}^* & 0
    \end{array}
    \right) 
    \left(
    \begin{array}{c}
    \hat{\psi}_{\mathbf{k}, 1} \\
    \hat{\psi}_{\mathbf{k}, 2}
    \end{array}
    \right), \label{effective-model}
\end{equation}
where $h_{\mathbf{k}} = t\sum_{i=1}^3 e^{i\mathbf{k} \cdot \Delta_i} $ with $t$ being the nearest neighbor hopping integral and $\hat{\psi}_{\mathbf{k},a}$ ($\hat{\psi}_{\mathbf{k},a}^\dagger$) is annihilation (creation) operator for the state $\psi_{\mathbf{k}, a}(\mathbf{r}) = \frac{1}{\sqrt{N_{\mathrm{UC}}}}\sum_{n} e^{i\mathbf{k}\cdot\mathbf{R}_{n,a} } w(\mathbf{r} - \mathbf{R}_{n, a} )$ on sublattice $a=1,2$; the sum is taken over all unit cells $\mathbf{R}_{n,a}$, and $N_{\mathrm{UC}}$ is the number of unit cells.
The eigenstates of $\tilde H_0$ reads 
\begin{equation}
    \psi_{\mathbf{k}, \pm} (\mathbf{r}) = \frac{1}{\sqrt{2}} ( g_{ \mathbf{k} } \psi_{\mathbf{k}, 1}(\mathbf{r}) \pm \psi_{\mathbf{k}, 2} (\mathbf{r}) ),\quad g_{ \mathbf{k} } = \frac{ h_{\mathbf{k}} }{ |h_{\mathbf{k}}| } 
\end{equation}
where $\pm$ denotes the eigenstates for conduction and valence bands, respectively.
Using the eigenstate basis and assuming that the Wannier function is well localized, i.e., $ \big< w(\mathbf{r}-\mathbf{R}_{n,a}) \big|e^{i\mathbf{q}\cdot \hat{\mathbf{r}} } \big| w(\mathbf{r}-\mathbf{R}_{m,b}) \big> \approx \delta_{nm} \delta_{ab} \big< w(\mathbf{r}-\mathbf{R}_{n,a}) \big|e^{i\mathbf{q}\cdot \hat{\mathbf{r}} } \big| w(\mathbf{r}-\mathbf{R}_{n,a}) \big>  $, the interaction matrix becomes that in Eq.~\eqref{eq:interaction-matrix-fit}.

The interaction matrix in Eq.~\eqref{eq:interaction-matrix-fit} corresponds to that for the honeycomb lattice model with a Gaussian Wannier function $w(\mathbf{r}) = \frac{1}{\sqrt{\pi}\xi_G} e^{-\mathbf{r}^2/2\xi_G^2}$, in which case $\eta_{\mathbf{q}}= \int \mathrm{d}^2 \mathbf{r} |w(\mathbf{r})|^2 e^{i\mathbf{q}\cdot \mathbf{r}}= e^{-\frac{q^2 \xi_G^2}{4} }$;
in this case, $\xi_G$ corresponds to the Wannier radius of the Gaussian orbital.
Hence, the large $q$ scattering channels, which include large-angle scatterings, are suppressed exponentially by increasing $\xi_G$.
We also performed a similar analysis using an exponentially-decaying Wannier function, which gives similar results as shown in the Supplemental Material~\cite{suppl}.

The honeycomb lattice model in Eq.~\eqref{effective-model} is one of the simplest model with the Dirac cones at $K$ and $K'$ points, while the nature of the interaction matrix is that of the normal model with a localized Wannier orbital.
For the electron-phonon interaction, a model similar to this model is known to reproduce the momentum dependence of the interaction matrix in moir\`e graphene~\cite{Ishizuka2021a}.
Hence, comparing our results to the phenomenological model in Eq.~\eqref{effective-model} highlights the characteristics of the interaction matrix.

Note that Eq.~\eqref{eq:interaction-matrix-fit} should be taken as a fitting function to elucidate the nature of the interaction matrix using a few parameters. 
Due to Wannier obstruction~\cite{Brouder2007a, Thouless1984a, Thonhauser2006a, Soluyanov2011a}, constructing a tight-binding model with exponentially localized Wannier functions requires considering high-energy orbitals, at least near the magic angles~\cite{Koshino2018a,Po2018a,Song2019a}.
Rather, we used Eq.~\eqref{eq:interaction-matrix-fit} as a fitting tool to compare to what extent the interaction matrix differs from typical cases.

\begin{figure}
    \centering
    \includegraphics[]{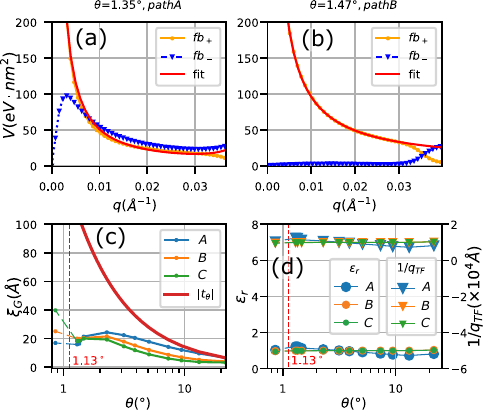}
    \caption{Fitting of the interaction matrix for the flat bands by using a Gaussian Wannier function. (a,b) The interaction matrix corresponds to scattering from the conduction to the conduction band (fb+), the interaction matrix corresponds to scattering from the conduction to the valence band (fb-), and the fitting function for fb+ (fit). The results are for paths $A$ (a) and $B$ (b). (c) Angle dependence of $\xi_G$ of Gaussian Wannier function calculated in three paths $A$, $B$, and $C$. The length of the primitive vector $|\mathbf{t}_\theta|$ is also plotted. (d)  The twist angle dependence of $\epsilon_r$ and $1/q_{\mathrm{TF}}$ for the three paths. See the main text for details.
    }\label{fig:gauss_angle_dependence}
\end{figure}

Figures~\ref{fig:gauss_angle_dependence}(a) and~\ref{fig:gauss_angle_dependence}(b) shows the interaction matrix of conduction bands $V_{ \mathbf{k}, \mathbf{k}', \mathbf{q} }^{3,4,3,4}$ for $\theta=1.35^\circ$ path $A$ and $\theta=1.47^\circ$ path $B$, respectively; they are calculated using the microscopic model (fb+) and a fitted interaction matrix using the effective tight-binding model (fit). The interaction matrix for the effective tight-binding model is obtained by fitting the results of the microscopic model using $\epsilon$, $q_{\mathrm{TF}}$, and $\xi_G$ as the fitting parameters. The interaction matrix is well fitted by that of the effective tight-binding model in a wide range of $q$, as in Fig.~\ref{fig:gauss_angle_dependence}(a) and Fig.~\ref{fig:gauss_angle_dependence}(b).
On the other hand, the sharp drop of $V_{ \mathbf{k}, \mathbf{k}', \mathbf{q} }^{3,4,3,4}$ near $q\sim 0.035$ \AA$^{-1}$, is not reproduced in the fitted function.
The discrepancy is presumably a consequence of an anticrossing of the conduction and valence bands in the microscopic model, as we show in~\cite{suppl}; such an anti-crossing is not present in the effective tight-binding model. 
Indeed, the interaction matrix between the conduction and valence bands, $V_{ \mathbf{k}, \mathbf{k}', \mathbf{q} }^{1,1,3,3}$, increases in the $q \ge 0.03$ \AA$^{-1}$, region as shown by the blue dots (fb-). 
Therefore, we fitted $V_{ \mathbf{k}, \mathbf{k}', \mathbf{q} }^{n,n',m,m'}$ excluding the large $q$ region where a band crossing or an anticrossing occurs.
With the treatment, the interaction matrix for the effective tight-binding model reproduces the overall behavior of $V_{ \mathbf{k}, \mathbf{k}', \mathbf{q} }^{n,n',m,m'}$.
However, the fitted parameter is $\xi_G=16.4$~\AA~ and $20.4$~\AA, which are considerably smaller than the lattice parameter $a\sim100$ \AA.
It implies that the large $q$ scattering is larger compared to what is naively expected from the Coulomb interaction.

Figures~\ref{fig:gauss_angle_dependence}(c) and~\ref{fig:gauss_angle_dependence}(d) show the twist angle dependence of the fitting parameters. As shown in Fig.~\ref{fig:gauss_angle_dependence}(d), relative permittivity $\epsilon_r$ and screening length $1/q_{\mathrm{TF}}$ are almost constant of the twist angle in the entire range of calculation.
On the other hand, $\xi_G$ shown in Fig.~\ref{fig:gauss_angle_dependence}(c) changes by changing the twist angle. In most materials, the radius of the Wannier orbital scales linearly with $|\mathbf{t}_{\theta}|$, in the case of which the dependence on the twist angle becomes $\xi_G\propto 1/\theta$ in the small $\theta$ limit. 
In contrast, the result in Fig.~\ref{fig:gauss_angle_dependence}(c) shows that $\xi_G\le |\bf t_\theta|/3$ at the angles $\theta\le 2^\circ$. 

For the regions $1.0^\circ \lesssim \theta\lesssim1.2^\circ$, on the other hand, we find that the interaction matrix shows a non-monotonic $q$ dependence which cannot be reproduced using the Gaussian model.
An example of the fitting is shown in the  Supplemental Material~\cite{suppl}.
One possible explanation for the failure might be the clover-leaf-like structure of the Wannier orbital~\cite{Koshino2018a}, which has many features not captured by the simple Wannier orbital.
Another possible cause of the failure is the Wannier obstruction~\cite{Brouder2007a, Thouless1984a, Thonhauser2006a, Soluyanov2011a}.
The Wannier obstruction prohibits constructing exponentially decaying Wannier orbitals. 
Recent theoretical studies find that the flat bands are topologically nontrivial, at least near the magic angles~\cite{Koshino2018a,Po2018a,Song2019a}.
As the obstruction prohibits constructing exponentially localized Wannier $q$ dependence, the Wannier obstruction may have non-trivial effect on $V_{ \mathbf{k}, \mathbf{k}', \mathbf{q} }^{n,n',m,m'}$.

In addition to the flat bands, we also calculated the interaction matrix for the high-energy bands. For them, the interaction matrix is well reproduced by the effective model, and $\xi_G$ is more than twice as large as that for the flat bands at the same twist angle at $\theta \lesssim 2.0^\circ$, as shown in the Supplemental Material~\cite{suppl} (see also~\cite{Shilov2024a} therein). Hence, comparing the transport and other properties of moir\'e graphene in the flat and high-energy bands may provide evidence for the anomalous interaction matrix of the flat bands near the magic angle.

\section{Conclusion}
In this work, we studied the electron-electron interaction matrix using a transferrable tight-binding model for carbon atoms and by constructing an effective tight-binding model. 
Using the transferrable tight-binding model, we find that the large-$q$ scattering by electron-electron interaction is enhanced in the flat bands.
For the twist angles $\theta\gtrsim1.2$, the enhancement of large-$q$ scattering is well reproduced by an effective compression of $\xi_G$; $\xi_G$ is smaller than the unit cell, reaching below $1/4$ of the length of primitive vectors at $\theta<2^\circ$.
The enhancement of large-$q$ scattering by a Wannier orbital with a small $\xi_G$ was predicted in a recent work~\cite{Ishizuka2022a}.
On the other hand, for $\theta\lesssim1.2^\circ$ we find a non-monotonic $q$ dependence of the interaction matrix.
Such a $q$ dependence cannot be explained by the effective tight-binding model. 
As the non-monotonic $q$ dependence of the interaction matrix is seen only at small twist angles near the first magic angle, it is possibly related to the unusual electronic properties of moir\'e graphene.
The enhanced electron-electron scattering taking place at $\theta\lesssim 2^\circ$, which is tunable by controlling the twist angle may provide another way of controlling the correlation effect.

The enhancement of large-$q$ scattering indicates that careful consideration of the interaction matrix might be necessary to understand the interaction physics of moir\'e graphene, in addition to the narrow bandwidth.
Considering that the electrons near the Fermi surface dominantly contribute to the transport, the enhanced large-$q$ scattering is likely to affect the transport properties of the moir\'e graphene.
Interestingly, such enhancement of the electron-electron scattering is seen below $\theta\sim4^\circ$.
This is different from the flat band formation, which is highly sensitive to the twist angle; the bandwidth is more than ten times larger than that of the magic angle at $\theta\sim1.1^\circ$.
Therefore, the twist-angle dependence of the material properties can be used to investigate the origin of the interaction physics in moir\'e graphene.

On the other hand, the unusual $q$ dependence of the interaction matrix near the magic angle was not reproducible within the phenomenological model considered.
One possible explanation for this could be the clover-leaf-like structure of the Wannier orbital~\cite{Koshino2018a}.
The form of the Wannier function significantly affects the interaction matrix in momentum space. The clover-leaf-like structure of the Wannier orbital may contribute to the unusual $q$ dependence. 
Another cause of failure might be the Wannier obstruction~\cite{Brouder2007a, Thouless1984a, Thonhauser2006a, Soluyanov2011a}.
The Wannier obstruction prohibits constructing exponentially decaying Wannier orbitals. As the $q$ dependence of the interaction matrix is affected by the functional form of the Wannier orbitals, the form of $V_{ \mathbf{k}, \mathbf{k}', \mathbf{q} }^{n,n',m,m'}$ may be significantly affected by the Wannier obstruction.
If such an effect exists, assuming the interaction matrix to be similar to that of the model in Eq.~\ref{effective-model} might be unsuitable. 
Therefore, a model incorporating the Wannier obstruction, such as those in Refs.~\cite{Song2022a, Calderon2020a} might provide a better starting point.
Regardless of the origin, the unusual $q$ dependence is likely to affect the low-temperature property of moir\'e graphene, through enhanced large $q$ scattering.
The results and analysis presented in this work unveil nontrivial modifications of the electron-electron scattering, which affect the correlation physics in moir\'e graphene.

\section*{Data availability}
The data supporting this study’s findings are available within the article and the Supplemental Material.

\begin{acknowledgments}
This work was supported by JSPS KAKENHI (Grants No. JP23K03275 and JP25H00841) and JST PRESTO (Grant No. JPMJPR2452).
\end{acknowledgments}

\bibliography{ref}

\end{document}


\preprint{APS/123-QED}

\title{
Supplemental Material to\\
``{\it  Anomalous enhancement of large-momentum scattering by electron-electron interaction in moir\'e superlattices}''
}

\author{Taiki Sato}
\affiliation{
Department of Physics, Tokyo Institute of Technology, Meguro, Tokyo, 152-8551, Japan
}

\author{Hiroaki Ishizuka}
\affiliation{
Department of Physics, Tokyo Institute of Technology, Meguro, Tokyo, 152-8551, Japan
}

\date{\today}


\pacs{
}

\maketitle
\onecolumngrid
\section{Tight-binding model}

The single-particle Hamiltonian is based on a transferrable tight-binding model for carbon atoms~\cite{Xu1992a, TdLaissardiere2010a},
\begin{equation}\begin{split}
    H_0 &= \sum_{n, \alpha, l, m, \beta, l'}  t(\mathbf{r}_{n, \alpha, l}-\mathbf{r}_{m, \beta, l'} )  \hat{c}_{n, \alpha, l}^\dagger \hat{c}_{m, \beta, l'}.
\end{split}\end{equation}
Here, $\mathbf{r}_{n, \alpha, l}$ is the position vector of sublattice $\alpha$ in the $n$th unit cell in the $l$th layer, $\hat{c}_{n, \alpha, l}$ ($\hat{c}_{n, \alpha, l}^\dagger$) the annihilation (creation) operator for the electron in the $p_z$ orbital of the atom at $\mathbf{r}_{n, \alpha, l}$, and
\begin{equation}\begin{split}
    t(\mathbf{r} ) = n_z(\mathbf{r})^2 t_{pp\sigma}(\mathbf{r}) + (1 - n_z(\mathbf{r})^2) t_{pp\pi} (\mathbf{r} ).\label{eq:hopping}
\end{split}\end{equation}
is the hopping integral between two $p_z$ orbitals at distance $\mathbf{r}=(x,y,z)$\cite{Slater1954a}.
In Eq.~\eqref{eq:hopping}, $n_z(\mathbf{r})=z/|\mathbf{r}|$ is the direction cosine for the $z$ axis, and 
\begin{equation}\begin{split}
    &t_{pp\pi}(\mathbf{r})= -\gamma_0 \exp\left( q_{\pi} \left(1 - \frac{\sqrt{3} |\mathbf{r}|}{a_0} \right)  \right),\\
    &t_{pp\sigma}(\mathbf{r})= \gamma_1 \exp\left( q_{\pi}\left(1 - \frac{|\mathbf{r}|}{d}\right)  \right).
\end{split}\end{equation}
are the $\pi$ and $\sigma$ hopping integrals between the $p_z$ orbitals, respectively~\cite{TdLaissardiere2010a}.
Here, $\gamma_0=2.7~\mathrm{eV}$ is the nearest neighbor hopping integral of graphene, $\gamma_1=0.48~\mathrm{eV}$ is the inter-layer hopping integral between two vertically aligned carbon atoms, and $d=3.35$ \AA~ is the distance between the two layers of graphene. In the following, we use $q_\pi=\ln(10)/(\sqrt{3}-1)$ and $q_\sigma = \sqrt{3}d q_\pi/a_0$, which reproduce the dispersion of pristine and bilayer graphene in the first-principles calculations~\cite{TdLaissardiere2012a}.

\section{exponential wannier function}

\begin{figure}[h]
    \centering
    \includegraphics[width=0.7\linewidth]{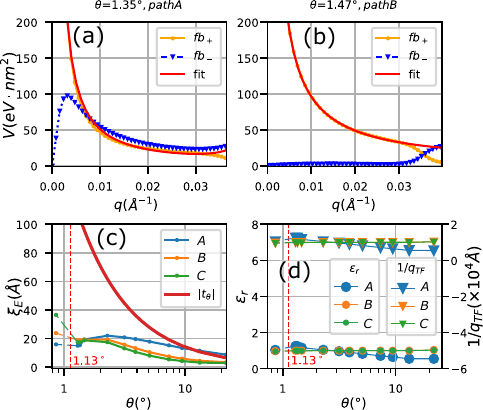}
    \caption{
    Fitting of the interaction matrix for flat bands by using an exponential Wannier function. (a,b) The interaction matrix corresponds to scattering from the conduction to conduction band (fb+), the interaction matrix corresponds to scattering from the conduction to valence band (fb-), and the fitting function for fb+ (fit). The results are for paths $A$ (a) and $B$ (b). (c) Angle dependence of radius $\xi_E$ of exponential Wannier function calculated in three paths $A$, $B$, and $C$. The length of the primitive vector $|\mathbf{t}_\theta|$ is also plotted. (d)  Angle dependence of $\epsilon_r$ and $1/q_{\mathrm{TF}}$ calculated in three paths. See the main text for details.
    }
    \label{fig:S_Fig1}
\end{figure}

Here, we considered an exponentially decaying Wannier function $w(\mathbf{r})= \sqrt{\frac{2}{\pi}} \frac{1}{\xi_E} e^{-r/\xi_E}$, where $\xi_E$ is the orbital radius.
Figure~\ref{fig:S_Fig1} shows an example of the fitting and the twist angle dependence of the fitted parameters. 
The results are similar to those of the Gaussian Wannier function.

\section{Electron-electron interaction matrix for high-energy bands}

\begin{figure}[h]
    \centering
    \includegraphics[width=0.7\linewidth]{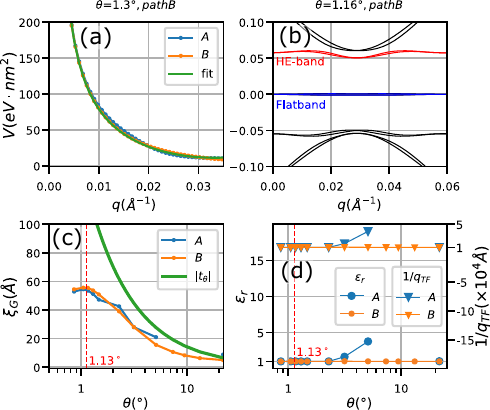}
    \caption{
    Interaction matrix and its twist-angle dependence of the high-energy bands. (a) The interaction matrix for paths $A$ and $B$, and the result of fitting (fit). (b) Band structure for the path $B$. The high-energy band is indicated by the red curve. (c) Angle dependence of radius $\xi_G$ of Wannier function calculated for two paths $A$ and $B$. The length of primitive vector $|\mathbf{t}_\theta|$ is also plotted. (d)  The angle dependence of $\epsilon_r$ and $1/q_{\mathrm{TF}}$ calculated for the two paths. Angles where the fitting parameters are strongly affected by band crossings are excluded. }
    \label{fig:S_Fig2}
\end{figure}

We also considered the interaction matrix for the high-energy dispersive bands shown in Fig.~\ref{fig:S_Fig2}(b).
Experimentally, the filling of moir\'e materials is highly tunable, enabling doping of the bands above the flat bands~\cite{Shilov2024a}. 
As the $\theta$ dependence of the interaction matrix in the high-energy band is generally different from that of the low-energy bands, comparing the results of the flat bands to those of the high-energy bands may provide a way to experimentally investigate the enhancement of el-el scattering in the flat bands.

The interaction matrix of the high-energy bands is calculated in the same way as those of flat bands. For the fitting, we considered a triangular lattice tight-binding model with Gaussian Wannier function $w(\mathbf{r})=\frac{1}{\sqrt{\pi} \xi_G} e^{-r^2/2\xi_G^2  }$ since the band bottom of the high-energy band is usually at the $\Gamma$ point and no Dirac nodes exist in the high-energy bands.
The eigenstate wavefunction of the high-energy band reads $\psi_{\mathbf{k}} (\mathbf{r})  = \frac{1}{\sqrt{N}}\sum_n e^{i\mathbf{k} \cdot \mathbf{R}_{n} }w(\mathbf{r}-\mathbf{R}_{n})$, where the sum is taken over all lattice points $\mathbf{R}_n$ in the triangle lattice. Using these results, the interaction matrix for high energy bands reads 
\begin{equation}
        V_{ \mathbf{k}, \mathbf{k}', \mathbf{q} }^{\mathrm{HE} } = \sum_{\overline{\mathbf{q}'} = \mathbf{q} } \tilde{V}(\mathbf{q}') |\eta_{\mathbf{q}'}|^2,
        \label{eq:highenergy-interaction-matrix-fit}
\end{equation}
which we use for the effective tight-binding model.

Figure~\ref{fig:S_Fig2}(a) shows the interaction matrix for $\theta=1.3^\circ$ and $V_{ \mathbf{k}, \mathbf{k}', \mathbf{q} }^{\mathrm{HE} }$.
Unlike the case of flat bands, $V_{ \mathbf{k}, \mathbf{k}', \mathbf{q} }^{\mathrm{HE} }$ shows a very small dependence on the paths.
In addition, the overall form of $V_{ \mathbf{k}, \mathbf{k}', \mathbf{q} }^{\mathrm{HE} }$ is well reproduced by the Coulomb interaction, except for angles where the fitting parameters were strongly affected by band crossings (see the section~\ref{sec:S3-B}).
Similar to the result shown in Fig.~\ref{fig:S_Fig2}(a), we find that the interaction matrix is well fitted by Eq.~\eqref{eq:highenergy-interaction-matrix-fit} for all twist angles we considered. 

Fig.~\ref{fig:S_Fig2}(c) and Fig.~\ref{fig:S_Fig2}(d) shows the twist angle dependence of the fitting parameters $\xi_G, \epsilon_r$, and $1/q_{\mathrm{TF}}$. 
 Similarly to the case of flat bands, $\epsilon_r$ and $1/q_{\mathrm{TF}}$ show very little dependence on the twist angle for the path $A$ and path $B$ below $2.5^\circ$.
 The upturn for the path $B$ in the $\theta\gtrsim2.5^\circ$ region is likely to be an artifact of the band crossings; we mainly focus on $\theta\lesssim2.5^\circ$ as the multi-band effect complicates the analysis. 
 On the other hand, $\xi_G$ increases as the twist angle decreases, contrasting the flat-band case; $\xi_G$ for the high-energy bands is more than twice as large as that for the flat bands at the same twist angle at $\theta\lesssim 2.0^\circ$.
 The larger $\xi_G$ suppresses large-$q$ scattering in the high-energy bands, suppressing the $T^2$ resistivity and other features related to the el-el interaction.
Hence, comparing the transport and other properties of moir\'e graphene in the flat and high-energy bands may provide evidence for the anomalous interaction matrix of the flat bands near the magic angle.

\section{Interaction matrix for the other bands}

\begin{figure}[b]
    \centering
    \includegraphics[]{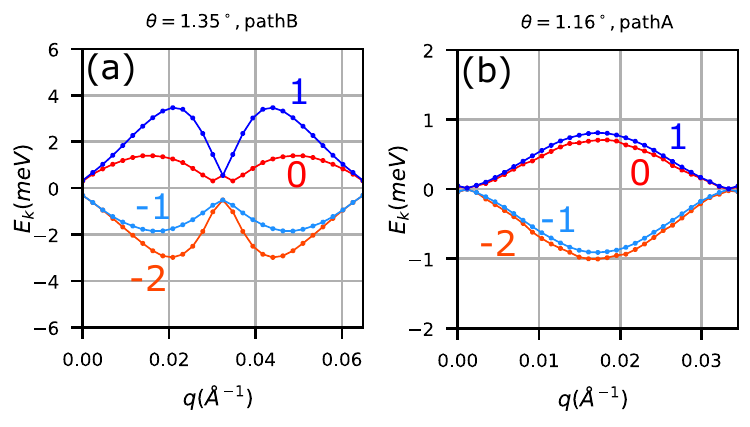}
    \caption{The flat bands calculated along the path B at twist angle $\theta = 1.35^\circ$ (a) and along the path A at twist angle $\theta=1.16^\circ$ (b). We labeled the bands in order of decreasing energy.
    }\label{fig:S_Fig3}
\end{figure}


Here, we use a slightly different notation from the main text. An example is shown in Fig.~\ref{fig:S_Fig3} (a), which shows the dispersion of the flat bands for $\theta=1.35^\circ$ along the path $B$ in the main text.
Here, we labeled the valence bands as $n=-2, -1$ and the conduction band as $n=0, 1$. The bands are labeled in the ascending order of energy. 

\subsection{Flat bands}
\begin{figure}
    \centering
    \includegraphics[]{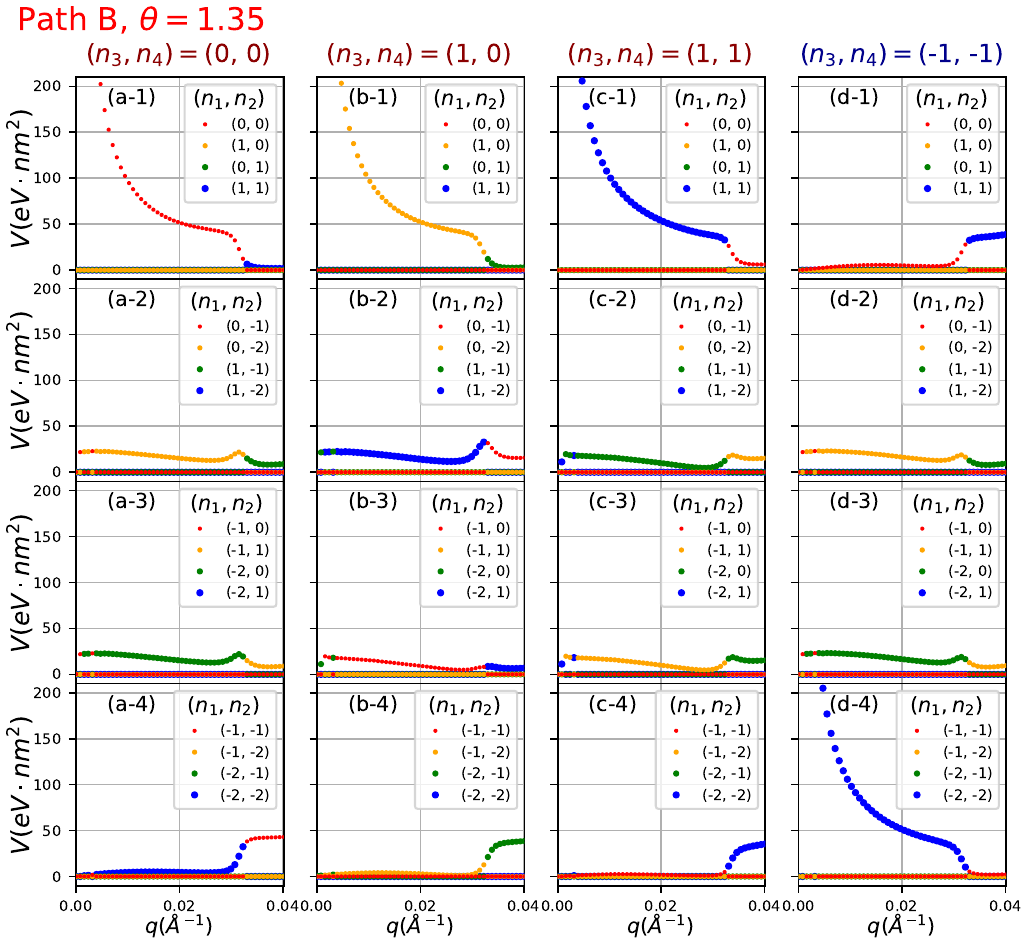}
    \caption{$|V_{ \mathbf{k}, \mathbf{k}', \mathbf{q} }^{n_1, n_2, n_3, n_4}|$ calculated along the path B at twist angle $\theta=1.35^\circ$. The chosen band indices are $(n_3, n_4)=(0, 0)$ in (a-1)-(a-4), $(n_3, n_4)=(1, 0)$ in (b-1)-(b-4), $(n_3, n_4)=(1, 1)$ in (c-1)-(c-4), and $(n_3, n_4)=(-1, -1)$ in (d-1)-(d-4). All combinations of the band indices $(n_1, n_2)$ chosen from the flat bands are plotted.
    }\label{fig:S_Fig4}
\end{figure}

Figures~\ref{fig:S_Fig4}(a-1)-\ref{fig:S_Fig4}(a-4) presents the interaction matrix $|V_{ \mathbf{k}, \mathbf{k}', \mathbf{q} }^{n_1, n_2, 0, 0}|$ for $(n_1, n_2)$ corresponding to the flat bands, calculated along the path $B$ at a twist angle $\theta = 1.35^\circ$; Fig.~\ref{fig:S_Fig4}(a-1) shows the amplitude of the scattering between the two conduction-band electrons and Fig.~\ref{fig:S_Fig4}(a-4) shows the amplitude of the scattering between conduction- and valence-band electrons. The intraband elements, $n_1=n_3$ and $n_2=n_4$, are essentially the only non-zero elements for $q < 0.03$ \AA, whereas the interband elements, $n_1\ne n_3$ and $n_2\ne n_4$, dominate for $q>0.03$ \AA. This behavior can be explained by an anticrossing between the conduction and valence bands, which is at $ q \sim 0.03 $ in Fig~\ref{fig:S_Fig3}. 

In addition, Fig.~\ref{fig:S_Fig4}(a-2) and \ref{fig:S_Fig4}(a-3) show the matrix elements that correspond to the amplitude of the scattering process in which one electron is scattered within the conduction bands and the other is scattered from the conduction bands to a valence band. In both cases, the values of $|V_{ \mathbf{k}, \mathbf{k}', \mathbf{q} }^{n_1, n_2, 0, 0}|$ lie between those shown in Fig.~\ref{fig:S_Fig4} (a-1) and (a-4).

Figures~\ref{fig:S_Fig4}(b-1)-\ref{fig:S_Fig4}(b-4) and Fig.~\ref{fig:S_Fig4}(c-1)-\ref{fig:S_Fig4}(c-4) presents the interaction matrix $|V_{ \mathbf{k}, \mathbf{k}', \mathbf{q} }^{n_1, n_2, n_3, n_4}|$ for $(n_3, n_4)$ corresponding to the conduction bands. Despite the differences in the band indices $(n_3, n_4)$, the $q$-dependence of the interaction matrix shows a behavior similar to those of the conduction bands, including the anti-crossing effect. For example, the intraband elements shown in Fig.~\ref{fig:S_Fig4} (a-1), (b-1), and (c-1) show similar characteristics.

Figures~\ref{fig:S_Fig4}(d-1)-\ref{fig:S_Fig4}(d-4) presents the interaction matrix $|V_{ \mathbf{k}, \mathbf{k}', \mathbf{q} }^{n_1, n_2, -1, -1}|$ for $(n_3, n_4)=(-1,-1)$, which corresponds to the valence bands. Here, Fig.~\ref{fig:S_Fig4}(d-4) corresponds to the amplitude of the scattering within the valence bands. Similar to the scattering amplitudes within the conduction bands, the intraband scattering element shown in Fig.~\ref{fig:S_Fig4}(d-4) is the largest for $q < 0.03$ \AA, whereas the interband scattering element shown in Fig.~\ref{fig:S_Fig4}(d-1) becomes prominent for $q>0.03$ \AA.

\begin{figure}
    \centering
    \includegraphics[]{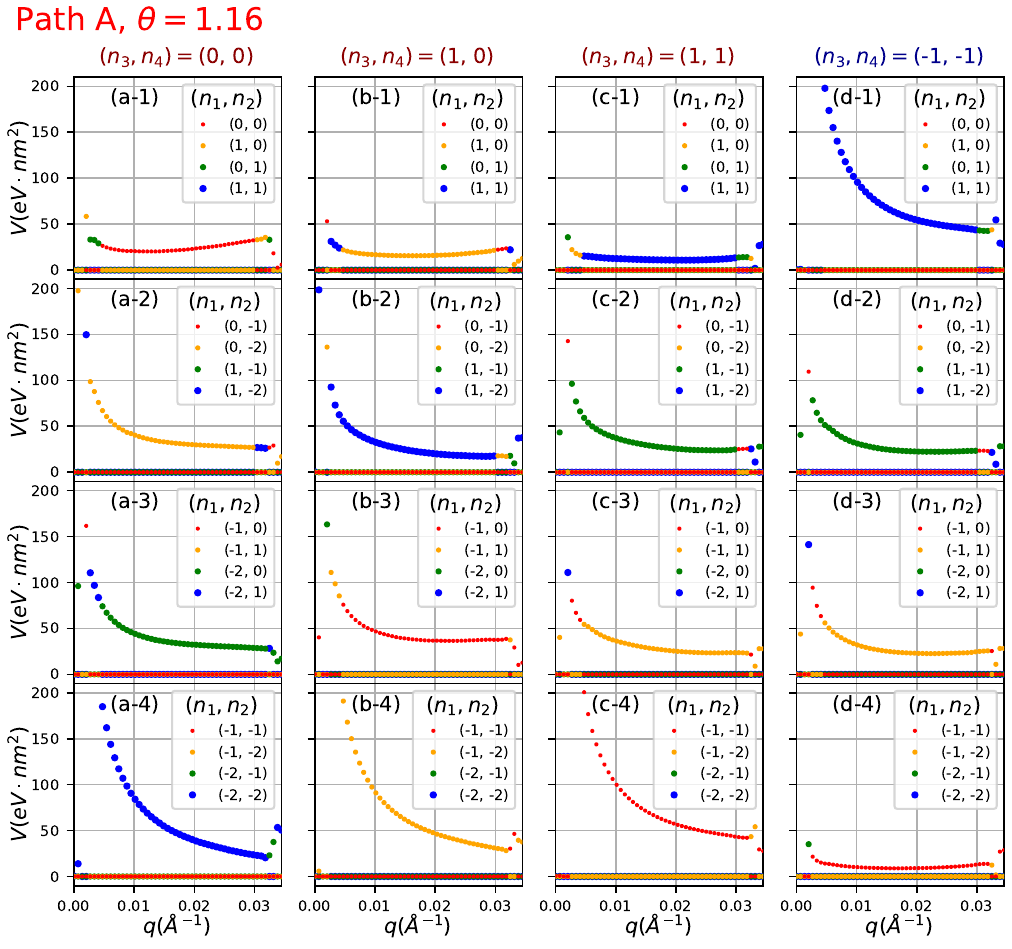}
    \caption{$|V_{ \mathbf{k}, \mathbf{k}', \mathbf{q} }^{n_1, n_2, n_3, n_4}|$ calculated along the path A at twist angle $\theta=1.16^\circ$. The chosen band indices are $(n_3, n_4)=(0, 0)$ in (a-1)-(a-4), $(n_3, n_4)=(1, 0)$ in (b-1)-(b-4), $(n_3, n_4)=(1, 1)$ in (c-1)-(c-4), and $(n_3, n_4)=(-1, -1)$ in (d-1)-(d-4). All combinations of the band indices $(n_1, n_2)$ chosen from the flat bands are plotted.
    }\label{fig:S_Fig5}
\end{figure}

Figure~\ref{fig:S_Fig5} shows the interaction matrix $|V_{ \mathbf{k}, \mathbf{k}', \mathbf{q} }^{n_1, n_2, n_3, n_4}|$ calculated along the path $A$ for twist angle $\theta = 1.16^\circ$. Similar to Fig.~\ref{fig:S_Fig4}, the interaction matrix shows a consistent pattern across different band indices. On the other hand, the interaction matrix in Figs.~\ref{fig:S_Fig5}(a-4), \ref{fig:S_Fig5}(b-4), \ref{fig:S_Fig5}(c-4), and \ref{fig:S_Fig5}(d-1) shows a large interband scattering amplitude. This is caused by the enhancement of the backscattering along the path $A$.

\subsection{\label{sec:S3-B}High-energy band}

\begin{figure}[h]
    \centering
    \includegraphics[]{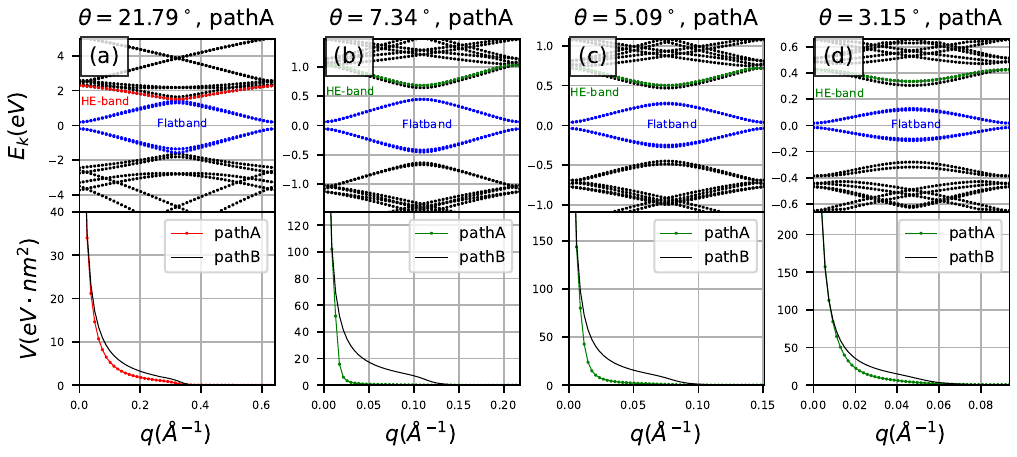}
    \caption{Band structure calculated along path A (top row) and the corresponding interaction matrix (bottom row). The twist angle is $\theta=21.79^\circ, 7.34^\circ, 5.09^\circ$ and $3.15^\circ$ for (a), (b), (c) and (d) respectively.
    }\label{fig:S_Fig6}
\end{figure}

The top rows of the Figure~\ref{fig:S_Fig6} display the band structure calculated along the path $A$ for twist angles (a) $\theta=21.79^\circ$, (b) $7.34^\circ$, (c) $5.09^\circ$, and (d) $3.15^\circ$. The bottom rows of the Figure~\ref{fig:S_Fig6} show corresponding interaction matrix $|V_{ \mathbf{k}, \mathbf{k}', \mathbf{q} }^{n_1, n_2, n_3, n_4}|$ calculated along the paths $A$ and $B$ for the high-energy band. 

The amplitude of the interaction matrix calculated along path $A$ for the twist angle $\theta=7.34^\circ$ and $5.09^\circ$ is much smaller than that calculated along path $B$. This is because the reconnection of the high-energy band occurs in these angles. For twist angle $\theta=21.79^\circ$, the high-energy band plotted by red (HE-band) does not intersect with other bands near $q=0$. However, for twist angle $\theta=7.34^\circ$, The high-energy bands are approaching bands of even higher energy, leading to an anti-crossing near $q=0.01$ \AA. Consequently,  the amplitude of the scattering process straddling the anti-crossing is suppressed. Because of band reconnection, the interaction matrix for the high-energy band calculated along path $A$ is much smaller than path $B$ and the fitting doesn't work. 

\section{Suppression of back scattering in the Dirac electrons}

 The amplitude $|V_{ \mathbf{k}, \mathbf{k}', \mathbf{q} }^{3,4,3,4}|$ for path A is smaller than those of paths B and C.
This suppression resembles the suppression of backscattering in the Dirac electrons. For pristine graphene, eigenstate of the tight-binding hamiltonian reads
\begin{equation}
    \psi_{\mathbf{k}, \pm} (\mathbf{r}) = \frac{1}{\sqrt{2}} ( g_{ \mathbf{k} } \psi_{\mathbf{k}, 1}(\mathbf{r}) \pm \psi_{\mathbf{k}, 2} (\mathbf{r}) ),\quad g_{ \mathbf{k} } = \frac{ h_{\mathbf{k}} }{ |h_{\mathbf{k}}| } 
\end{equation}
where $h_{\mathbf{k}} = t\sum_{i=1}^3 e^{i\mathbf{k} \cdot \mathbf{b}_i} $ with $t$ being the nearest neighbor hopping, $\mathbf{b}_i$ is vector connecting the three nearest neighbor sites in honeycomb lattice of pristine graphene and $\psi_{\mathbf{k}, a}(\mathbf{r}) = \frac{1}{\sqrt{N_{\mathrm{UC}}}}\sum_{n} e^{i\mathbf{k}\cdot\mathbf{R}_{n,a} } p_z(\mathbf{r} - \mathbf{R}_{n, a} )$ with sublattice $a=1,2$; the sum is over all unit cells $\mathbf{R}_{n,a}$, and $N_{\mathrm{UC}}$ is the number of unit cells, $p_z(\mathbf{r})$ is the $p_z$ orbital of the atom.

The amplitude depends on the factor $(g_{\mathbf{k}_1}^*g_{\mathbf{k}_2} + 1) $.
For states near the Dirac point $\mathbf{K} =  (\frac{4\pi}{3a}, 0)$, using $\tilde{\mathbf{k}}=\mathbf{k}-\mathbf{K}$, $ g_{ \mathbf{k} } \approx - e^{i\varphi_{ \tilde{\mathbf{k}} }}$, where $\varphi_{ \tilde{\mathbf{k}} }=\arg( \tilde{k}_x + i \tilde{k}_y )$ .
Thus, the factor reads $ (g_{\mathbf{k}_1}g_{\mathbf{k}_2} + 1) \approx  \big( e^{i ( \varphi_{ \tilde{\mathbf{k}}_2 } -  \varphi_{ \tilde{\mathbf{k}}_1 } ) }  + 1 \big)$. If two momenta $\mathbf{k}_1=\mathbf{k-q}$, and $\mathbf{k}_2=\mathbf{k}$ are positioned at opposite sides of the Dirac point, $(g_{\mathbf{k}_1}g_{\mathbf{k}_2} + 1) \approx 0$ and the amplitude $|V_{ \mathbf{k}, \mathbf{k}', \mathbf{q} }^{nmn'm'}|$ is suppressed. This is the suppression of the backscattering in the Dirac electrons.

\section{Fitting of interaction matrix near the magic angle}

\begin{figure}[h]
    \centering
    \includegraphics[]{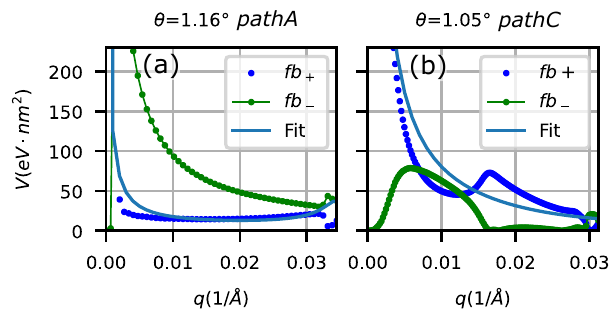}
    \caption{Fitting of the interaction matrix for the flat bands by using a Gaussian Wannier function. (a,b) The interaction matrix corresponds to scattering from the conduction to the conduction band (fb+), the interaction matrix corresponds to scattering from the conduction to the valence band (fb-), and the fitting function for fb+ (fit). The results are for paths $A$ (a) and $C$ (b).
    }\label{fig:S_Fig7}
\end{figure}

Figures~\ref{fig:S_Fig7}(a) and~\ref{fig:S_Fig7}(b) shows the interaction matrix of the conduction bands $V_{ \mathbf{k}, \mathbf{k}', \mathbf{q} }^{3,4,3,4}$ at angles near the magic angle, $\theta=1.16^\circ$ and $\theta=1.05^\circ$, calculated using the microscopic model (fb+) and a fitted interaction matrix using the effective tight-binding model (fit). The interaction matrix for the effective tight-binding model is obtained by fitting the results of the microscopic model using $\epsilon$, $q_{\mathrm{TF}}$, and $\xi_G$ as the fitting parameters. The effective tight-binding model failed to reproduce the interaction matrix calculated by the microscopic model, especially for the range of small $q$. Also, the angle dependence of some of the fitting parameters show anomalous values around the magic angle. In addition, the fitting parameters considerably vary depending on the path. For example, at the twist angle $\theta = 1.16^\circ$, the radius is $\xi_G\sim0.7$ \AA for the path $A$, while it is $\xi_G\sim20$ for the paths $B$ and $C$. The inconsistencies suggest a failure of the fitting.

\section{Calculation for the path that connects the M point to the $\Gamma$ point}

\begin{figure}[h]
    \centering
    \includegraphics[width=0.7\linewidth]{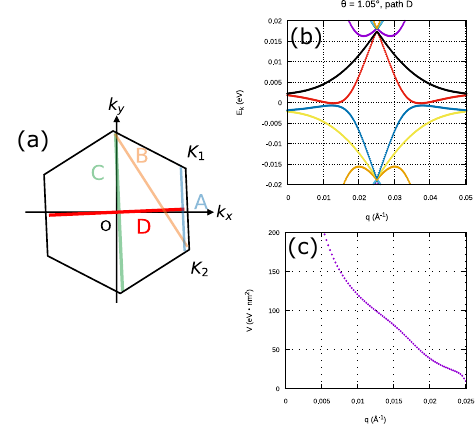}
    \caption{(a) Four paths along which the interaction matrix was calculated. (b) The band structure for a twist angle of $1.05^\circ$ along path D. (c) The interaction matrix for twist angles $\theta=1.05^\circ$ along path D.
    }\label{fig:S_Fig8}
\end{figure}

Figure~\ref{fig:S_Fig8}(c) presents the interaction matrix for a twist angle of $\theta=1.05^\circ$ along path D, which is illustrated in Fig.\ref{fig:S_Fig8}(a). The bands along this path is shown in the Fig.\ref{fig:S_Fig8}(b). The behavior is similar to that of path B, and no unusual behavior is noted.

%
%
%
\bibliography{ref} 